\documentclass[aps,superscriptaddress,prb,twocolumn,floatfix]{revtex4}

\usepackage{graphicx}
\usepackage{graphics}
\usepackage{subfigure}
\par

\def\bea{\begin{eqnarray}}
\def\eea{\end{eqnarray}}
\def\ben{\begin{equation}}
\def\een{\end{equation}}
\def\benu{\begin{enumerate}}
\def\enu{\end{enumerate}}

\def\br{{\bf r}}

\begin{document}

\title{Same-spin dynamical correlation effects on the electron localization}

\author{Stefano Pittalis}
\affiliation{Center S3, CNR Institute of Nanoscience, Via Campi 213/A, 41125 Modena, Italy.}
\author{Alain Delgado}
\affiliation{Center S3, CNR Institute of Nanoscience, Via Campi 213/A, 41125 Modena, Italy.}
\affiliation{Centro de Aplicaciones Tecnol\'ogicas y Desarrollo Nuclear, Calle 30 \# 502, 11300 La Habana, Cuba.}
\author{Carlo Andrea Rozzi}
\affiliation{Center S3, CNR Institute of Nanoscience, Via Campi 213/A, 41125 Modena, Italy.}

\maketitle

\subsection*{Abstract}
 
The Electron Localization Function (ELF) -- as proposed  originally by Becke and Edgecombe -- has been widely adopted as
a descriptor of atomic shells and covalent bonds. 
The ELF takes into  account the  antisymmetry of Fermions but it  neglects
the multi-reference character of a truly interacting many-electron state.
Electron-electron interactions induce, schematically, different kind of correlations:
non-dynamical correlations mostly affect stretched molecules and strongly correlated systems;
dynamical correlations dominate in weakly correlated systems. Here,
within an affordable computational effort, we estimate the effects of 
same-spin dynamical correlations on the electron localization by means of a simple modification of the ELF.


\section{Introduction}

At the core of quantum chemistry, there is the concept of chemical bond.
Since early times, there have been ongoing efforts to set up effective descriptors of bonding in molecules and solids.
Pair of electrons of opposite spins were proposed as the  basis of bonding in the Lewis theory~\cite{Lewis}.
More recently the idea emerged that  bonding and other structure, such as atomic shells, should reflect the ``localization" of electrons in the system.

An effective way to estimate the degree of localization is to exploit the fact that, if the position of one electron is fixed at a given point in space, it must be 
less likely to find a second electron around the same position. If the pair is made of electrons with parallel spin states, the
localization should be related to the properties of the Fermi hole function which reflects the effect of Pauli exchange 
repulsion~\cite{Fermihole}.

Becke and Edgecombe have followed these lines of thoughts and have  analyzed a state of the form of a single Slater determinant~\cite{BE}.
In this case, the {\em conditional} distribution $P^{\sigma}_{\rm cond}(\br_1,\br_2)$ 
of having a $\sigma$-spin electron at position $\br_2$ given that
there is a $\sigma$-spin electron at the reference position $\br_1$ is given by
\ben\label{p2}
P^{\sigma}_{\rm cond}(\br_1,\br_2) = \rho_\sigma(\br_2) + h^{\sigma}_{\rm x}(\br_1,\br_2)\;
\een
(we  employ a simplified notation with a single spin index as the pair of electrons is restricted to parallel spin state), 
where the exchange-hole (x-hole) function
\ben
h^{\sigma}_{\rm X}(\br_1,\br_2) := - \frac{ | \rho^{\sigma}_1(\br_1,\br_2) |^2 }{ \rho_\sigma(\br_1) }\;
\een
is expressed through the spin-dependent one-body reduced density matrix
\ben\label{rho1}
\rho^{\sigma}_1(\br_1,\br_2) = \Sigma_{i} \psi_{i\sigma}^*(\br_1)\psi_{i\sigma}(\br_2)\;.
\een
Here, the summation is restricted to occupied $\sigma$-state single-particle orbitals, $\psi_{i \sigma}(\br)$,  and
the particle density is readily obtained as $\rho_\sigma(\br)=\rho^{\sigma}_1( \br, \br)$.

Among the fundamental features of the x-hole function, there are:
\begin{itemize}
\item its normalization
\ben
\int d^3r_2~ h^{\sigma}_{\rm X}(\br_1,\br_2) = -1\;,
\een
i.e, it accounts for $-1$ electron;

\item its average extent
\ben\label{R}
R^{\sigma}_{\rm X}(\br) :=  \left[ \int  \frac{ | h^{\sigma}_{\rm x}(\br,u)| }{u} du \right]^{-1} \;
\een 
where
\ben
h^{\sigma}_{\rm x}(\br,u) := \frac{1}{4 \pi } \int d \Omega h^{\sigma}_{\rm x}(\br, \br + {\bf u})
\een
is the spherical average of the x-hole function determined around the reference position;

\item its short-range behavior for small $u$ 
\bea\label{ahx}
h^{\sigma }_{\rm X}(\br,u) &=& -\rho_\sigma(\br) - \frac{1}{6} \left[ \nabla^2\rho_\sigma(\br) - 2D_\sigma(\br) \right]  u^2  \nonumber 
\\  &+& \cdot\cdot\cdot
\eea
where the first term is also known as the on-top x-hole and the coefficient of the second term is the curvature of the x-hole. In detail 
\ben
D_\sigma(\br) := \left[ \tau_\sigma(\br) - \frac{1}{4} \frac{ \left( \nabla \rho_\sigma(\br) \right)^2}{\rho_\sigma(\br) } \right]\;,
\een
with
\ben
 \tau_\sigma(\br)  := \Sigma_{i} | \nabla \psi_{i\sigma}(\br) |^2
\een
being the (double) of the (positively defined) kinetic energy density.
\end{itemize}
Similarly to Eq.~(\ref{ahx}),
\ben\label{apd}
\rho_\sigma(\br,u) = \rho_\sigma(\br) + \frac{1}{6} \nabla^2 \rho_\sigma(\br) u^2 + \cdot \cdot \cdot\;
\een 
and insertion of both Eq.~(\ref{ahx}) and Eq.~(\ref{apd}) in Eq.~(\ref{p2}) yields
\ben\label{key_b}
P^{\sigma}_{\rm cond}(\br,u) = \frac{1}{3} D_\sigma(\br) u^2 + \cdot\cdot\cdot\;.
\een
Note, Eq.~(\ref{key_b}) is a key expression that we will later modify [see Eq.~(\ref{cP})] to take into account correlations beyond exchange effects
to some extent.

For an inhomogenous system, $D_\sigma(\br)$ exhibits a non-trivial dependence on the
reference position $\br$. It turns out useful to make a comparison with respect to a {\em local} uniform reference
with same density as the actual system. Becke and Edgecombe introduced the quantity
\ben
\chi_\sigma(\br) = D_\sigma(\br) / D^{\rm unif}_\sigma(\br)
\een
where 
\ben
D^{\rm unif}_\sigma(\br) = \frac{3}{5} \left( 6 \pi^2 \right)^{2/3} \rho^{5/3}_\sigma(\br) 
\een
from which they defined the Electron Localization Function (ELF) as
\ben\label{ELF}
\mathrm{ELF}(\br) = \frac{1}{ 1 + \chi^2_\sigma(\br) }\;.
\een 
The relevant information provided by the ELF is contained in its variations.
The function varies from zero to one: for ELF = 0.5, the electrons are localized as in the uniform reference, 
otherwise they are more (less) $\mathrm{ELF} > 0.5$
($\mathrm{ELF} <  0.5$) localized. 
The modulation of the ELF nicely reproduces the shell structure in atoms and
well emphasizes covalent molecular bonds in molecules~\cite{Savinreview}. 

The ELF  discussed above is  built from a  wavefunction of {\em non-interacting} electrons.
Works in the literature have  proposed generalizations of the ELF~\cite{Silvi2010,Kohout1,Kohout2} at the full correlated level.
Here, we find it useful to reconsider this topic by distinguishing different type of correlations.
While non-dynamical correlations  affect stretched molecules and strongly correlated systems, 
dynamical correlations  dominate in  most of the molecules at typical equilibrium distances and weakly correlated systems.
In the language of multi-reference methods, non-dynamical correlations are related to the mixing of configurations of low energy (which may be degenerate or almost degenerate),
while dynamical correlations have to do with the mixing of  configurations at higher energy. 
Can one estimate the effects of  {\em dynamical} correlations on the electron localization?
Can these estimates be made within a well-affordable computational approach?
As we illustrate in the next sections, these objectives can be achieved by exploiting the ability of density functional 
approximations to capture semilocal correlations that are of dynamical type.
The approach introduced here
neglects  opposite-spin contributions, not because they are expected to be unimportant but because we are
concerned with an estimation implying a simple and direct modification of the original ELF.

Finally, we point out that the study of electron localization has  been fundamental for the understanding of how to capture 
non-trivial non-localities in the exchange-correlation functional by means of simple approximations.
This and related topics continuously  attract attention and  novel insights have been discussed in recent  contributions~\cite{jSun13,Ent,RG1,RG2}.
The analysis presented here adds useful information.

The work is organized as follows: 
in Sec.~\ref{newELF}, we present  a  modified ELF;
in Sec.~\ref{test}, we analyze results of applications to atoms, molecules, and Jellium;  in the Sec.~\ref{co}, we discuss conclusions and outlooks.

\section{Modification of the ELF}\label{newELF}

In an interacting system, we must account for the fact that electrons tend to avoid each other not only because of the antisymmetry
requirement but also because of the electron-electron repulsion. As mentioned in the introduction, we shall focus on  same-spin pairs,
for which the conditional distribution is given by
\ben\label{p2c}
P^{\sigma}_{\rm cond}(\br_1,\br_2) = \rho_\sigma(\br_2) + h^{\sigma }_{\rm x}(\br_1,\br_2) + h^{\sigma }_{\rm c}(\br_1,\br_2)
\een
where $h^{\sigma}_{\rm c}(\br_1,\br_2)$ describes many-body corrections beyond the effects included in the
(Generalized) Kohn-Sham [(G)KS] state (notice that this definition differs from the coupling-constant averaged correlation hole).

For a pair of electrons at small inter-particle distance, the interaction is strong yet the relative kinetic term is not suppressed. 
Thus, the pair in this configuration looks like a ``hydrogen atom" but with a repulsive interaction and much lighter nucleous~\cite{cusp}. 
Density functional theory (DFT) has successfully contributed to clarify that
semilocal density functional approximations are very effective to capture such short-ranged correlations -- that are, in fact, of dynamical type.

Our choice is to resort to the model proposed (and explained in full details) by Becke~\cite{cB88}  
for the spherical average of $h^{\sigma }_{\rm c}(\br_1,\br_2)$ for $\br_2 \approx \br_1$. 
Becke's final target was the DFT hole, whereas we are interested in the hole at full coupling strength:
\ben\label{hmodel}
h^{\sigma }_{c}(\br,u) = \frac{u^2 \left[ u-z_\sigma(\br) \right] D_\sigma(\br)}{6 \left( 1 + z_{\sigma}(\br)/2 \right)} F(\gamma_\sigma(\br) u)\;,
\een
where
\ben\label{z}
z_\sigma(\br) := 2 c R^{\sigma}_{\rm X}(\br)\;,
\een
has the meaning of  a ``correlation length" (it sets the shortest inter-particle distance at which the correlation hole vanishes at each reference position).  
In the following, we shall employ $c = 0.88$ as determined by Becke by fitting DFT correlation energies. The stability of the results with respect to 
relatively large variation of $c$  will be demonstrated in the next section. $F(x)$ is a damping function with quadratic small-$x$ behavior  ($F(x) \sim 1 - ax^2 + ...$).
$\gamma_\sigma(\br)$ will not enter in our final expression but  it is  instructive to appreciate that, it is determined by enforcing  proper 
zero-normalization condition of the correlation hole and as a result,  $\gamma_\sigma(\br) \approx 1/z_\sigma(\br)$.
For small $u$, Eq.~(\ref{hmodel}), reads 
\ben\label{sr-hmodel}
h^{\sigma}_{c}(\br,u) = - \frac{z_\sigma(\br) D_\sigma(\br)}{6\left( 1 + z_\sigma(\br)/2 \right)} u^2\; +\cdot\cdot\cdot \;.
\een
The model in Eq.~(\ref{hmodel}) has several appealing features: it sets a natural correlation length; it contains  zero electron as the corresponding exchange-hole
 accounts for $-1$ electrons; it gives vanishing correlation energy for one electron systems; 
and its short-range behavior is expected to be realistic particularly for inhomogenous systems.

Upon insertion of Eq.~(\ref{sr-hmodel}), Eq.~(\ref{apd}), and Eq.~(\ref{ahx})  into (the spherical average  around ${\bf r}_1 = {\bf r}$ of) Eq.~(\ref{p2c}), we get
\ben\label{cP}
P^{\sigma}_{\rm cond}(\br,u) \approx \frac{1}{3}\left\{ D_\sigma({\bf r}) \left[ 1- \frac{z_\sigma(\br)}{2\left( 1 + z_\sigma(\br)/2 \right)} \right]  \right\} u^2\; +\cdot\cdot\cdot \;
\een
from which we define a ``modified" ELF (mELF) as follows
\ben\label{cELF}
\mathrm{mELF}(\br) := \frac{1}{1 +  \chi'^{2}_\sigma(\br)}
\een
where
\bea\label{chiprime}
\chi'_\sigma(\br) &=& \chi_\sigma(\br) 
 \left[ 1 - \frac{z_\sigma(\br)}{2\left( 1 + z_\sigma(\br)/2 \right)} \right] \nonumber \\
&=& \frac{  \chi_\sigma(\br)  }{ 1 + z_\sigma(\br) /2 }\;.
\eea
Notice that the numerator of Eq.~(\ref{chiprime}) includes
the reference to the uniform gas as originally defined by Becke and  Edgecombe.

Finally, also notice that Eq.~(\ref{cELF}) -- through Eq.~(\ref{chiprime}), (\ref{z}) and Eq.~(\ref{R}) -- entails the calculation of the Slater potential  $U^{\sigma}_{\rm X}(\br)$ 
\ben\label{Ux}
U^{\sigma}_{\rm X}(\br) = - \frac{1}{R^{\sigma}_{\rm X}(\br)}\;;
\een
i.e., the Slater potential is the potential due to the  exchange-hole function. 
Features and implications of the modified ELF are explored in Sec.~\ref{test}.

\subsection{Further simplifications}
Since an evaluation of the Slater potential, $U^{\sigma}_{\rm X}({\br})$, may not be readily available in some DFT codes,
we remind that this quantity can be  approximated in terms of the Becke-Russel (BR) model $U^{\sigma}_{\rm BR89, X}(\br)$~\cite{BR89} .
The BR potential is appealing for finite systems because it entails a model for the x-hole which
is exact for finite $1$-particle system and reproduces exactly  the shot-range behavior of $h^{\sigma}_{\rm X}(\br,u)$ in Eq.~(\ref{ahx})
of  any  $N$-electron system. One should keep in mind that, by construction, it produces a potential of a ``localized" hole function.
Within our scope, the BR model can be straightforwardly applied using orbitals obtained from a converged (G)KS calculation.
 
In the case of molecules, where the exchange-hole can delocalize over  multi centers, 
the option to adopt the full Slater potential is  available from direct post processing of 
the  orbitals obtained from a converged  (G)KS calculation.

For applications in extended metallic-like systems,  we may  consider the LDA expression for the Slater potential
\ben\label{UxLDA}
U^{{\rm LDA},\sigma}_{\rm X}(\br)= - 3\left[ \frac{3}{4 \pi} \rho_\sigma(\br)  \right]^{1/3}\;.
\een
This expression is exact for Jellium.

\subsection{Current-carrying and time-dependent states}
In the expressions above, we have tacitly assumed real-valued orbitals.
For current-carrying states in the case of ground state degeneracy ~\cite{jD, jB, jT, jP} and
time-dependent processes~\cite{jELF},
complex-valued orbitals must be considered instead. 
The proper expressions are obtained with the substitution 
$
\tau(\br) \rightarrow \tau(\br) - \left( \frac{ {\bf j}_p(\br) }{\rho(\br)} \right)^2\;,
$
where ${\bf j}_p(\br)$ is the paramagnetic current.
Of course, a time-dependence $(\br) \rightarrow (\br,t)$  must also be admitted for the time-dependent case. 
In the reminder of this work, we shall restrict ourselves to systems with closed-shell ground states.

\section{Applications and further analyses}\label{test}

In this section, we apply  Eq.~(\ref{ELF}) and Eq.~(\ref{cELF}) to several systems in order to quantify their differences.

\subsection{Atoms}
\begin{figure}[htb]
\includegraphics[width=0.9\columnwidth]{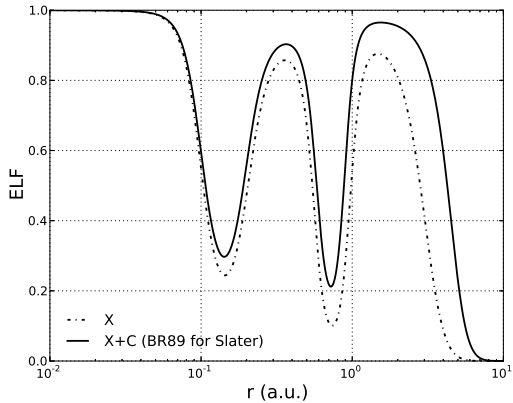}
\caption{(X) Electron Localization Function in its original form, Eq.~(\ref{ELF}), and (X+C) in its  modified version, Eq.~(\ref{cELF}), computed for the Argon atom.
Kohn-Sham orbitals obtained with the APE code~\cite{APE}, using a LDA for the exchange-correlation energy functional,
are used as input to the two expressions. The Slater potential, which enters Eq.~(\ref{cELF}) through the correlation length $z_\sigma$ [see also Eq.~(\ref{z}) and Eq.~(\ref{Ux})], is evaluated within the BR approximation (see also Fig.~\ref{fig2}).}
\label{fig1}
\vskip -0.5cm
\end{figure}

As a first system, we consider the Argon atom.
Fig.~\ref{fig1} reports the ELF and its  modified version  (i.e, mELF).
Qualitatively, the resulting pictures of the shell structure exhibits strong similarities.
The included short-ranged correlations  do not change the position of the maxima and minima.
However, we can  appreciate  quantitative differences:  minima are less shallow and  maxima higher.
Localization is enhanced and it increases and extents as moving outwardly from the center of the atom.

\begin{figure}[htb]
\includegraphics[width=0.9\columnwidth]{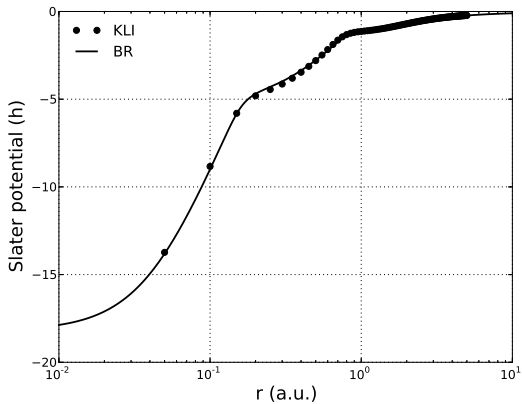}
\caption{Dots show the Slater potential for the Argon atom evaluated within the OEP-KLI method \cite{OEP} 
as implemented in the OCTOPUS code~\cite{octopus} (a grid with equally spaced points is employed). 
The solid line shows the same quantity evaluated within the BR approximation (this was then
employed in the calculation of the modified ELF reported in Fig.~\ref{fig1}). 
The BR potential is evaluated on quantities produced with the APE code as explained
in the caption of Fig.~\ref{fig1}. It is apparent that, in the region of interest, the two potentials do not differ significantly.
}
\label{fig2}
\vskip -0.5cm
\end{figure}

Fig.~\ref{fig1} was obtained by employing the BR model for the Slater potential.
In order to make the illustration self-contained, we report the BR and Slater potential for the Ar atom in 
Fig.~\ref{fig2}.

\begin{figure}[htb]
\includegraphics[width=0.9\columnwidth]{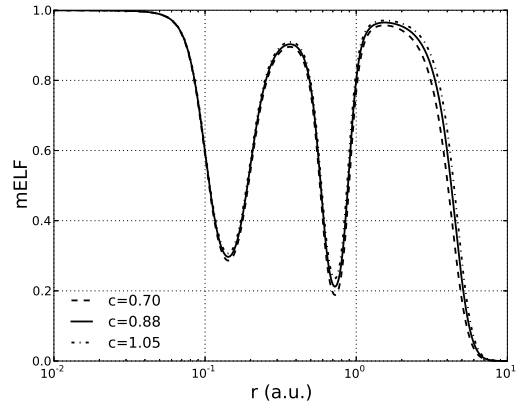}
\caption{Demonstration of the stability of the results reported in Fig.~\ref{fig1} for the modified ELF [ Eq.~(\ref{cELF})].   
Here, the parameter $c$ -- entering the final expression through Eq.~(\ref{z}) and Eq.~(\ref{chiprime}) --
is varied $\pm 20\%$
around its prescribed valued.}
\label{fig3}
\vskip -0.5cm
\end{figure}

Finally, the dependence of  the results on the values of the parameter $c$ of Eq.~(\ref{z}) is studied in Fig.~\ref{fig3}.
It is apparent that variations of about $\pm 20\%$ do not change the results significantly. 
Hence, we do not find any evidence of the necessity to  optimize the value of $c$ further.

\subsection{Molecules}
In order to assess the effect on the ELF of dynamic correlation, as introduced in this work, we have compared the usual ELF with  the mELF 
for a few small hydrocarbons with different CC bond orders (Ethane, Ethene and Ethyne). 

We have verified that the pictures obtained with the standard ELF is essentially unchanged with respect to what reported, for instance, in Ref.~\cite{Savinreview}.
The corrections introduced by mELF are only minor in these cases. More noticeable differences may be found in molecules involving heavier atoms.  

\begin{figure}[h]
\includegraphics[width=0.9\columnwidth]{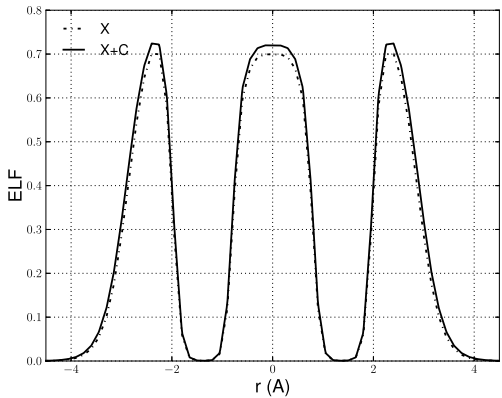}
\caption{(X) Electron Localization Function in its original form, Eq.~(\ref{ELF}), and (X+C) in its modified version, Eq.~(\ref{cELF}), for the Iodine molecule. 
The calculation was carried out with the OCTOPUS code using a LDA for the exchange-correlation energy functional. In (X+C), we have employed the Slater potential obtained 
{\em post}-LDA. We employed  Troullier-Martins norm-conserving pseudo potentials (as the focus here is on the state of the valence electrons).}
\label{fig4}
\vskip -0.5cm
\end{figure}

Let us consider the case of Iodine molecule in Fig.~\ref{fig4}. 
The qualitative behavior of mELF follows that of the ELF with larger values in the bonding region. We have verified that, 
when the two atoms are brought closer and closer, the relative minimum at the midpoint of both ELF and mELF is turned into a maximum when the bonding distance is reached (approximately 2.7 \AA). At this point, the electrons pairing from the two 5p atomic orbitals becomes a $\pi$ type molecular orbital. 
For  all the molecules considered in this section, we have input the {\rm mELF} with the  Slater potential evaluated with converged LDA KS-DFT results.

\subsection{Jellium}

The application of Eq.~(\ref{cELF}) to Jellium (an extended uniform gas of interacting electrons, whose charge is balanced by 
a smeared positive background) gives us the opportunity to put forward some explorative speculations.

For this system, the original {\rm ELF} is a constant {\em independent} of the particle density (the values of the constant being $0.5$).
Instead, our {\rm mELF} gives a constant that depends on the values of the particle density. 
Eq.~(\ref{z}) together with Eq.~(\ref{Ux}) and Eq.~(\ref{UxLDA}) show that the correlation length is in inverse relation with the particle density
$z_\sigma(\br) \sim \rho^{-1/3}_\sigma(\br) $.

For Jellium, the low-density limit corresponds to the strongly interacting regime.
If we require the KS wave functions to be plane waves, we find that the correlation length  diverges $z_\sigma(\br) \rightarrow +\infty$  and thus ${\rm mELF} \rightarrow 1$.
Since it is believed that, in the same limit, the electrons in Jellium tend to localize into a crystalline structure (i.e., the Wigner's crystal)~\cite{GV05},  
we may say that, {\rm mELF} captures this instability.

In the opposite limit, electrons in Jellium get weakly interacting. 
Correspondingly  $z_\sigma(\br) \rightarrow 0$ and thus  ${\rm mELF} \rightarrow  0.5$.
We may say that, electrons fully delocalize in order to get packed together to develop the high-density limit.

\section{Conclusions and outlooks}\label{co}
We have  shown how to account for the effects of same-spin dynamical correlations on the electron localization
within a model which requires an affordable computational effort.
We could visually asses  that the effects of these correlations are 
somewhat unimportant in small organic molecules. They imply some noticeable effects for atomic 
shells and bonds involving relatively heavier atoms. However, qualitative features such as the positions of the shells 
and the bond character obtained at the exchange-only level are unchanged. 
As a novel interesting feature, the proposed modified electron localization function 
appears to connect the degree of localization in Jellium to different interaction regimes.
For the future, it is appealing to attempt to include information on the opposite-spin channels
and of non-dynamical correlation effects.\\

{\it Acknowledgments --} This work was financially supported by the European Community through the FP7's MC-IIF MODENADYNA, grant agreement No. 623413.


\begin{thebibliography}{10}

\bibitem{Lewis}
G.N. Lewis. The atom and the molecule. {\it Journal of the American Chemical Society}, 38:762--785, 1916.


\bibitem{Fermihole}
R.F.W. Bader and M.E. Stephens. Spatial localization of the electronic pair and number distributions in molecules. {\it Journal of
the American Chemical Society}, 97:7391--7399, 1975; R.F.W. Bader, R.J. Gillespie and P.J. MacDougall, {\em ibid.} 110:7329, 1988.

\bibitem{BE}
A.D. Becke and K. Edgecombe. A simple measure of electron localization in atomic and molecular systems. {\it Journal of Chemical
Physics}, 92(9):5397--5403, 1990.

\bibitem{Savinreview} 
A. Savin, R. Nesper, S. Wengert and T.E. F{\"a}ssler. ELF: The Electron Localization Function. {\it Angewandte Chemie International
Edition}, 36:1808--1832, 1997.

\bibitem{Silvi2010} 
F. Feixas, E. Matito, M. Duran, M. Sol{\'a} and B. Silvi. Electron Localization Function at the Correlated Level: A Natural
Orbital Formulation. {\it Journal of Chemical Theory and Computation}, 6:2736--2742, 2010.

\bibitem{Kohout1} 
M. Kohout, K. Pernel, F.R. Wagner, Yu. Grin. Electron localizability indicator for correlated wavefunctions. I. Parallel-spin
pairs.{\it Theoretical Chemistry Accounts}, 112(5), 453--459, 2004.

\bibitem{Kohout2} 
M. Kohout, K. Pernel, F.R. Wagner, Yu. Grin. Electron localizability indicator for correlated wavefunctions. II Antiparallel-spin
pairs. {\it Theoretical Chemistry Accounts}, 113(5):287--293, 2005.


\bibitem{jSun13}
J. Sun, B. Xiao, Y. Fang, R. Haunschild, P. Hao, A. Ruzsinszky, G. I. Csonka, G. E. Scuseria, and J. P. Perdew, Phys. Rev. Lett. 111: 106401--106401-5, 2013.

\bibitem{Ent}
S. Pittalis, F. Troiani, C.A. Rozzi, and G. Vignale. Ab initio theory of spin entanglement in atoms and molecules.
{\it Physical Review B}, 91(7):075109, 2015.

\bibitem{RG1}
M.J.P. Hodgson, J.D. Ramsden, T.R. Durrant and R.W. Godby. Role of electron localization in density functionals. {\it Physical
Review B}, 90(24):241107(R), 2014.

\bibitem{RG2}
T.R. Durrant, M.J.P. Hodgson, J.D. Ramsden, R.W. Godby. Electron localization in static and time-dependent systems.
arXiv:1505.07687, 2015.



\bibitem{cusp}
R.T. Pack and W.B. Brown, {\it Journal of Chemical Physics}, 45:556-559, 1966.
 

\bibitem{cB88}
A.D. Becke. Correlation energy of an inhomogeneous electron gas: A coordinate‐space model. {\it Journal of Chemical Physics},
88(2):1053--1062, 1987.

\bibitem{BR89}
A.D. Becke. Exchange holes in inhomogeneous systems: A coordinate-space model. {\it Physical Review A}, 39(8):3761--3767, 1988.

\bibitem{jELF}
T. Burnus, M.A.L. Marques and E.K.U. Gross. Time-dependent electron localization function. {\it Physical Review A},
71(1):010501(R), 2005.

\bibitem{jD}
J. Dobson. Alternative expressions for the Fermi hole curvature. {\it Journal of Chemical Physics}, 98(11):8870--8872, 1993.

\bibitem{jB}
A.D. Becke. Current-density dependent exchange-correlation functionals. {\it Canadian Journal of Chemistry}, 74(6):995--997, 1996.

\bibitem{jT} J. Tao. Explicit inclusion of paramagnetic current density in the exchange-correlation functionals of current-density
functional theory. {\it Physical Review B}, 71(20):205107, 2005.

\bibitem{jP}
S. Pittalis, E. R\"as\"anen and E.K.U. Gross. Gaussian approximations for the exchange-energy functional of current-carrying
states: Applications to two-dimensional systems. {\it Physical Review A}, 80(3):032515, 2009.

\bibitem{APE} M. Oliveira and F. Nogueira. Generating relativistic pseudo-potentials with explicit incorporation of semi-core
states using APE, the Atomic Pseudo-potentials Engine. {\it Computer Physics Communications}, 178(7):524--534, 2008.

\bibitem{OEP}
T. Grabo, T. Kreibich, S. Kurth, and {E.K.U. Gross}. {S}trong
 {C}oulomb {C}orre\-lations in {E}lectronic {S}tructure {C}alculations:
 {B}eyond {L}ocal {D}ensity {A}pproxi\-mations. In V. Anisimov Gordon
 and Breach (Eds.), pp. 203, Amsterdam, 2000. 

\bibitem{octopus} A. Castro, H. Appel, M. Oliveira, C.A. Rozzi, X. Andrade, F. Lorenzen, M.A.L. Marques, E.K.U. Gross and
A. Rubio. Octopus: a tool for the application of time-dependent density functional theory. {\it Physica Status Solidi B},
243(11):2465--2488, 2006.

\bibitem{GV05}
G.F. Giuliani and G. Vignale. Quantum Theory of the Electron Liquid. {\it Cambridge University Press}, 2005.

\end{thebibliography}
\end{document}